# Integrating superconducting van der Waals materials on paper substrates

*Jon Azpeitia[a], Riccardo Frisenda[a]\*, Martin Lee[b], Damian Bouwmeester[b], Wenliang Zhang[a], Federico Mompean[a], Herre S. J. van der Zant[b], Mar García-Hernández[a]\*, Andres Castellanos-Gomez[a]\**

[a.] *Materials Science Factory. Instituto de Ciencia de Materiales de Madrid (ICMM-CSIC), Madrid, E-28049, Spain.*
[b.] *Kavli Institute of Nanoscience, Delft University of Technology, Lorentzweg 1, Delft, The Netherlands.*

Paper has the potential to dramatically reduce the cost of electronic components. In fact, paper is 10 000 times cheaper than crystalline silicon, motivating the research to integrate electronic materials on paper substrates. Among the different electronic materials, van der Waals materials are attracting the interest of the scientific community working on paper-based electronics because of the combination of high electrical performance and mechanical flexibility. Up to now, different methods have been developed to pattern conducting, semiconducting and insulating van der Waals materials on paper but the integration of superconductors remains elusive. Here, the deposition of $NbSe_2$, an illustrative van der Waals superconductor, on standard copy paper is demonstrated. The deposited $NbSe_2$ films on paper display superconducting properties (e.g. observation of Meissner effect and resistance drop to zero-resistance state when cooled down below its critical temperature) similar to those of bulk $NbSe_2$.

## Introduction

The combination between low cost (~ 0.1 €/m2), biodegradability and flexibility makes paper-based electronics very promising for applications like disposable wearable electronics and sensors.[1–10] However, the integration of novel electronic materials, like the van der Waals (vdW) materials family, on paper substrates is hampered by their fibrous structure that introduces a large surface roughness and liquid-absorption. For these reasons, standard device fabrication approaches, developed and optimized for the fabrication of devices on silicon substrates, cannot be directly used on paper substrates. New fabrication techniques have been developed in the last years to overcome that issue, and now vdW materials can be deposited onto paper substrates using inkjet printing of inks prepared by liquid-phase exfoliation [7,11–19] or by a recently reported all-dry abrasion-induced deposition method.[20,21] Up to now these methods already demonstrated that one can fabricate heterostructures and pattern complex devices, with high spatial resolution, with conducting, semiconducting and insulating vdW materials on paper.[13–16] During the elaboration of this manuscript Novoselov, Lu and co-workers demonstrated the preparation of inks of monolayer $NbSe_2$ through a mild electrochemical exfoliation method and their use to print superconducting films on $SiO_2$/Si substrates by inkjet printing.[22] The integration of vdW superconductors on paper substrates, however, is still lacking and its experimental realization is the goal of this manuscript.

**Experimental**

We deposited NbSe$_2$, as an illustrative example of vdW superconductor, on paper substrates by simply rubbing NbSe$_2$ fine powder (≥ 99.8%, average particle size of 5 µm, Alfa Aesar PN: 13101.09) against the surface of a piece of standard copy paper with a cotton swab. During the rubbing process the flakes are subjected to friction forces that abrades the vdW platelet crystals exfoliating them and leading to the deposition of a dense network of interconnected flakes.[20,21] The abrasion-induced deposition is carried out at atmospheric conditions until forming a homogeneous dark gray/black film. At this point we test the electrical continuity of the film with a handheld multimeter. A good electrically continuous film typically yields a resistance below 1 kΩ between two probes separated ~1-2 mm. We have determined the sheet resistance of two films yielding values between 100 Ω/□ and 300 Ω/□ (See the Supporting Information Figure S1). We address the reader to Ref. [[23]] for more detailed discussion about the resistivity measurements of abrasion-deposited van der Waals materials on paper.

Figure 1a shows pictures acquired during the deposition process of a NbSe$_2$ film on paper. A desktop vinyl cutter (Portrait Silhouette) has been used to fabricate a custom-made stencil mask that allows one to control the geometry of the deposited films with accuracy. By slicing the film with a sharp razor blade, we can even image the cross-section to estimate the average thickness of the film (see Figure 1b) which is typically around 20 µm.

**Results and discussion**

We have characterized the morphology and the chemical composition of the deposited NbSe$_2$ film on paper by scanning electron microscopy (SEM) and energy dispersive X-ray spectroscopy (EDX) using a FEI Helios G4 CX system. Figure 1c show SEM images with low (top panel) and high (bottom panel) magnification of the NbSe$_2$ film on paper. The NbSe$_2$ is deposited over the whole surface of the paper except for few spots of uncoated or barely coated paper that show up in the low magnification SEM image as brighter spots (highlighted with yellow arrows in Figure 1c top panel). From the SEM images we infer that abrasion during the deposition process crushes the NbSe$_2$ flakes forming a rather compact film. Inside the gaps between the paper fibers, however, it is sometimes possible to resolve NbSe$_2$ flakes with lateral sizes in the 1–5 µm range (slightly smaller than lateral dimensions of the flakes in the source NbSe$_2$ powder, see Figure S2). EDX spectroscopy analysis provides an insight about the chemical composition of the as-drawn NbSe$_2$ film. Figure 1d compares EDX spectra acquired on the source NbSe$_2$ powder, on the NbSe$_2$ deposited filling in the gaps between paper fibers and on the NbSe$_2$ deposited onto the outermost paper fibers. While the EDX spectra of NbSe$_2$ powder and NbSe$_2$ deposited between fibers are comparable, the spectra acquired on top of the outermost paper fibers show signals of NbSe$_2$ together with signals of oxygen, carbon and calcium originating from the paper underneath. In fact, A calcium carbonate (a bright white mineral with chemical formula $CaCO_3$) is quite often added to paper pulp as a filler. This result indicates that the NbSe$_2$ film on top of the outermost paper fibers is rather thin. See Figures S2 to S4 in the Supporting Information for further SEM and EDX analysis.

We further characterized the morphology of the samples by X-ray diffraction (XRD) to clarify if the NbSe$_2$ flakes have a preferential orientation induced by the deposition method. Figure 1e compares the X-ray diffraction (XRD) patterns, acquired with Cu Kα1 radiation, on the source NbSe$_2$ powder, on the NbSe$_2$ film on paper and on the bare (uncoated) paper. The XRD pattern of NbSe$_2$ powder agrees well with the expected for bulk NbSe$_2$. The XRD pattern of bare paper allows us to identify the features originated from the substrate in the XRD pattern measured on the NbSe$_2$-on-paper film. From a direct comparison between the patterns acquired on the powder and on the film on paper one can see how on the film the (00n) diffraction peaks are systematically more intense than the other peaks, indicating a preferential alignment of the NbSe$_2$ flakes with their basal plane parallel to the paper surface.

In order to characterize the superconducting properties of the as-deposited NbSe$_2$ films we have first used a superconducting quantum interference device (SQUID) magnetometer (Quantum Design) equipped with a 5 Tesla coil. **Figure 2a** shows the measured magnetization as a function of the temperature after a zero-field cooling, using a measuring field of 10 Oe in a direction parallel to the substrate and film. A clear transition is observed around $T$ ~7 K that can be interpreted as the onset of the superconducting state exhibiting a strong and sudden decrease of the magnetization due to the Meissner effect. Another evidence of the superconducting transition comes from the isotherm curve measured at 1.8 K in the inset: the NbSe$_2$-on-paper sample was cooled in zero-field from 300 K to 1.8 K and a magnetization *versus* magnetic field curve, starting from zero field, was measured. The inset in Figure 2a shows the diamagnetic response of the NbSe$_2$ sample on paper after subtraction of the paper contribution. Note the diamond like shape of the cycle pointing out to the typical magnetic response of a superconductor below its critical temperature $T_c$. **Figure 2b** shows a close-up of the isothermal first-magnetization curves for various temperatures. Linear fits to the curves at low field render the first critical field values $H_{c1}$ of a typical type II superconductor (shown in the inset of Figure 2b). $H_{c2}$ is larger than the 5 Tesla accessible with our SQUID coil.

We have also studied the temperature dependence of the resistance of a ~5×5 mm$^2$ NbSe$_2$ film on paper with the van der Pauw configuration in a physical property measurement system (PPMS, by Quantum Design) cryostat equipped with a 9 Tesla coil (with the magnetic field perpendicular to the film and paper substrate). **Figure 3** shows the resulting resistance *vs*. temperature ($R$ *vs*. $T$ hereafter) where there is an overall increase of the resistance upon sample cooldown. This comes to a surprise since metallic character has been reported for NbSe$_2$ and thus one would expect a continuous decrease of the resistance when decreasing the temperature. We attribute this behavior to localization due to the disordered nature of the film which can be described in terms of a thermally activated hopping mechanism between the interconnected flakes of the film as detailed later. Several thermally activated transport mechanisms, such as Arrhenius-like or variable range hopping, have yielded similar temperature dependences in other percolative systems composed of highly-conductive particles with a highly-resistive particle-to-particle interface.[24,25]

No specific feature linked to the onset of a charge density wave transition, with a reported transition temperature around $T = 145$ K and 35 K for single-layer and bulk samples respectively, has been observed in Figue 3.[26,27] However, several jumps of the resistivity while decreasing the temperature are apparent. Interestingly, at $T \sim 6$ K the *R vs. T* changes the overall trend and the resistance drops rapidly upon cooldown. This turning point matches well with the critical temperature observed in the SQUID measurement and points to the transition of the $NbSe_2$ flakes into the superconducting state. The high degree of disorder in the film and the interflake hopping resistance, however, prevents from observing a sharp electrical transition to a zero-resistance state. In fact, here we recall the polycrystalline nature of our superconducting film, with a broad distribution of flakes with various thickness, which could exhibit different critical temperatures,[28–33] and a variety of lateral sizes. Note that the large residual resistance below the superconducting transition is expected due to the large film channel length (~5 mm), orders of magnitude larger than the typical $NbSe_2$ platelets size, as the current has to pass through several interflake junctions that behave as a normal-state resistor. We anticipate that shorter channel devices will present a sharper transition with a reduced residual resistance. In fact recent results, reported during the elaboration of this work, on inkjet printed films of interconnected $NbSe_2$ layers on $SiO_2/Si$ substrates show a very abrupt superconducting transition with negligible residual resistance with electrode separation of <20 μm.[22] Inkjet printing, however, cannot be used for printing devices on standard copy paper, since the low-viscosity nanoinks tend to leak through the cellulose fibers.

The observed behaviour in the film of $NbSe_2$ flakes on paper can be qualitatively explained by a model of a random resistor network of interconnected $NbSe_2$ flakes. A small network of 4×4 resistors is shown, as a simplified example, in **Figure 4a**. The network is composed of two different kind of resistors called SC and I, which respectively represent the resistance of the superconducting $NbSe_2$ flakes (SC) and the flake-to-flake hopping (I) as depicted in Figure 4a. To populate the random resistor network, we define a probability parameter $p$ ($0 < p < 1$), which controls the percentage of I and SC resistors in the network. For each resistor we generate a random number $x$ uniformly distributed between 0 and 1. If $x < p$ we assign the resistor to the state I and if $x \geq p$ to SC. The simulations presented in the main text are performed using $p = 0.65$, a value that ensures a configuration with small clusters of SC resistors surrounded by I resistors. This configuration emulates the $NbSe_2$ platelets (SC islands) connected through flake-to-flake junctions (I resistors), see **Figure 4a**. The network can be biased by applying a voltage to the nodes in the network, in the 4×4 example the left side is kept at 1 V (red circles) and the right side at 0 V (black circles). After solving for the unkonwn voltages (white circles) one can calculate the current flowing through each resistor and the total current in the network using Ohm's law. See Supporting Information Figures S5 to S7 for more details about the random resistor network model and for simulations with different critical temperature and probability $p$ parameters.

**Figure 4b** shows the temperature dependence of the SC and I resistances used for the simulation. We have considered a variation of the critical temperature $T_c$ of the SC resistors from 1 K to 7 K to account for the thickness dependent $T_c$ of NbSe$_2$.[28–33] The I resistors are modelled as a thermally activated resistors. **Figure 4c** shows the mean resistance calculated from 10 different 120×120 random resistors networks as a function of the temperature, which shows a decrease of resistance at ~6 K without dropping all the way to zero ohms, a behaviour similar to that observed experimentally. The observation of a residual resistance below $T_c$ can be explained by the presence of the inter-platelet normal-state resistances, which are connected in series to the superconducting elements and thus limit the supercurrent flow. The model also reproduces the overall resistance increase upon cooling down for temperatures above $T_c$. The insets show the current magnitude maps calculated in the normal (20 K) and the superconducting (3 K) state. At 3 K higly conductive filaments (red filaments in the inset in Figure 4c) arise because of the transition of the SC resistors to the low resistance state, reducing the mean resistance of the network. Note that no filaments fully composed of SC resistors bridge the two electrodes. Our model also predicts that shorter channel devices will present more abrupt superconducting transitions and with a lower residual resistance below the superconducting transition, as discussed above. See the Supporting Information Figure S8 for the results of simulations of resistor networks with shorter and shorter channel lengths where this trend is shown.

We have further studied the magnetic field dependence of the NbSe$_2$ film on paper resistance (e.g., the inset in Figure 3 shows the magnetic field dependence of the *R vs. T* curve around $T_c$). As expected for a superconducting transition, as the magnetic field increases the critical temperature is depleted to a point such that above 3 Tesla the transition is basically supressed in the explored temperature range. **Figure 5a** shows the magnetic field dependence of the NbSe$_2$ film on paper resistance (*R vs. H*) at different temperatures. Above $T_c$ the system exhibits a negative magnetorresistance and the resistance decreases as an external magnetic field is applied. Below $T_c$, the magnetoresistance changes sign and the resistance increases dramatically when the magnetic field is applied. This behavior can be rationalized in terms of the model proposed by Porat *et al.* for highly disordered superconductors.[34] Again, the rationale behind is that a highly disordered superconductor can be viewed as a set of superconducting percolating paths. As a magnetic field is applied the smaller or less robust superconducting flakes enterinto the normal state and the number of superconducting paths decreasses and thus the resistance increases. As the field is further increased, the conduction evolves to a point where the superconductor paths are not connected and thus the normal current paths become dominant. The inset in Figure 5a shows the applied magnetic field when the the normal state sets in for the measured temperatures. This could be interpreted as the temperature dependence of an effective upper critical field $H_{c2}$ of the film.

We have also measured the angle dependence of the magnetoresistance of another NbSe$_2$ film on paper, with a bar-shape (5×2 mm$^2$ and probed with four-terminal sensing geometry. Comparing resistance measurements at two-terminal and four-terminal sensing we can infer that the contact resistance ranges from ~10 kΩ at room temperature to ~500 kΩ at ~ 4K. We used another PPMS (Quantum Design) cryostat equipped with a 14 Tesla coil (with the magnetic field perpendicular to the film and paper substrate) and a rotator mount for the sample. **Figure 5b** shows the magnetic field dependence of the NbSe$_2$ film resistance, at 5 K, for several angles ranging from -90º and 90º. The angle is labelled as -90º and 90º for magnetic fields perpendicular to the film, $H_\perp$, and as 0º for magnetic field parallel to the film, $H_\parallel$. The inset in Figure 5b plots the upper critical field $H_{c2}$ values extracted for different magnetic field tilting angles, displaying a huge anisotropy. Similar anisotropy has been also observed in single crystal NbSe$_2$, in agreement with the preferential alignment of the NbSe$_2$ platelets with their basal plane parallel to the paper surface (see the XRD data in Figure 1c).[33]

To gain a deeper insight into the electrical transport characteristics of the NbSe$_2$ films on paper below the superconducting critical temperature we have performed further magnetotransport measurements with a dilution fridge in the 100 mK to 1000 mK temperature range. Voltage *vs.* current curves have been measured at different magnetic fields (perpendicular to the film) on another NbSe$_2$ film on paper device (~5×5 mm$^2$ with van der Pauw geometry, see Figure 6a). For fields below 7 T, the curves display a zero-resistance state indicating fully developed superconducting transport across the sample. We have found that the electrical characteristics of the film do not change substantially in the studied temperature range. Figure 6b shows the derivative of the voltage *vs.* current traces (d$V$/d$I$) which facilitate the observation of the zero-resistance states at low current biasing. Figure 6c summarizes the d$V$/d$I$ as a function of the magnetic field and the current bias conditions through a false color map where it is shown how the superconductivity is completely suppressed by magnetic fields higher than ~6.5 T.

**Conclusions**

In summary, we show a facile route to integrate van der Waals superconductors on standard copy paper substrates by simply rubbing powder of the selected superconductor against paper. We illustrate it with NbSe$_2$, a prototypical vdW superconductor, finding that the as-deposited NbSe$_2$ films expel magnetic fields at temperatures lower than 7 K, a clear proof that they display Meissner effect. The resistance of NbSe$_2$ on paper also shows an abrupt drop at temperature lower than 6 K, consistent with a superconducting transition of the NbSe$_2$ flakes composing the interconnected network of platelets that is the film. This behavior can be accurately reproduced with a random resistor network model. Furthermore, magnetotransport measurements carried out at 150 mK show a complete superconducting transition, displaying a zero-resistance state across the sample (at millimeters scale), with a critical field of ~6.5 T. The results shown here are robust (all the 5 studied NbSe$_2$-on-paper samples present a superconducting transition) and general and thus they open the door to integrate other vdW superconductors on standard paper substrates. We believe that, given the low cost and low weight of paper substrates, this technique can become a new route towards mass-scalable production of simple superconducting devices (like superconducting high frequency filters) or as coatings for magnetic field shielding in cryogenic applications.

**Conflicts of interest**

**Acknowledgements**

This project has received funding from the European Research Council (ERC) under the European Union's Horizon 2020 research and innovation program (grant agreement n° 755655, ERC-StG 2017 project 2D-TOPSENSE) and the European Union's Horizon 2020 research and innovation program under the Graphene Flagship (grant agreement number 785219, GrapheneCore2 project and grant agreement number 881603, GrapheneCore3 project). R.F. acknowledges the support from the Spanish Ministry of Economy, Industry and Competitiveness (MINECO) through a Juan de la Cierva-formación fellowship 2017 FJCI-2017-32919 and the grant MAT2017-87134-C2-2-R. This work was supported by the Netherlands Organisation for Scientific Research (NWO/OCW), as part of the Frontiers of Nanoscience program.

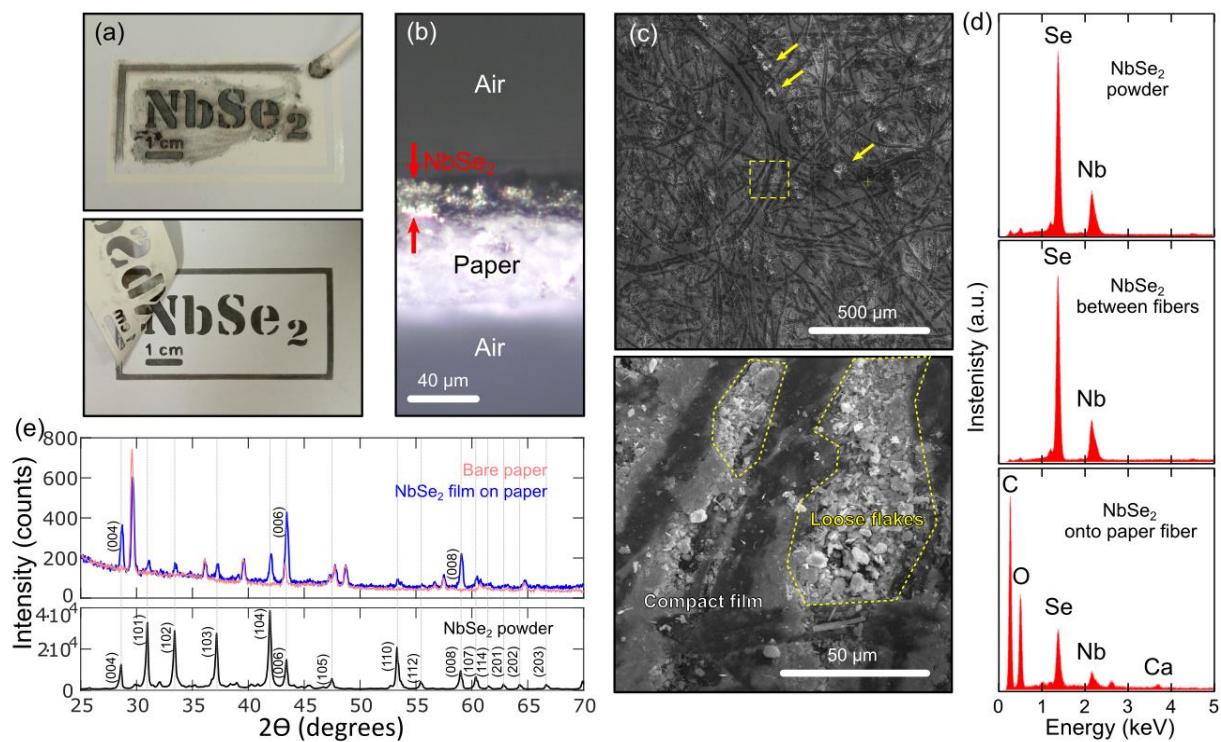

**Figure 1**. **Deposition of a NbSe$_2$ film on standard copy paper.** (a) Pictures acquired during the deposition process of a NbSe$_2$ film on paper following a user-defined pattern thanks to a vinyl stencil mask. (b) Optical microscopy image of a cross-section of the paper coated with a NbSe$_2$ film where the thickness of the NbSe$_2$ can be estimated. (c) SEM images showing the morphology of the NbSe$_2$ film on paper. The bottom panel shows a zoomed-in image of the region highlighted with a dashed yellow square in the top panel, displaying the different morphologies between the material deposited onto the outermost paper fibers and that deposited filling in the gaps between fibers layered structures and flakes that fill in the gaps between the paper fibers. (d) EDX analysis of the chemical composition of the source NbSe$_2$ powder material, the NbSe$_2$ deposited in the gap between the paper fibers and that deposited onto the outermost paper fibers. In the spectrum acquired on the NbSe$_2$ film deposited onto the paper fibers the signal of oxygen, carbon and calcium, originated by the paper substrate, arises pointing towards a lower thickness of the NbSe$_2$ film.

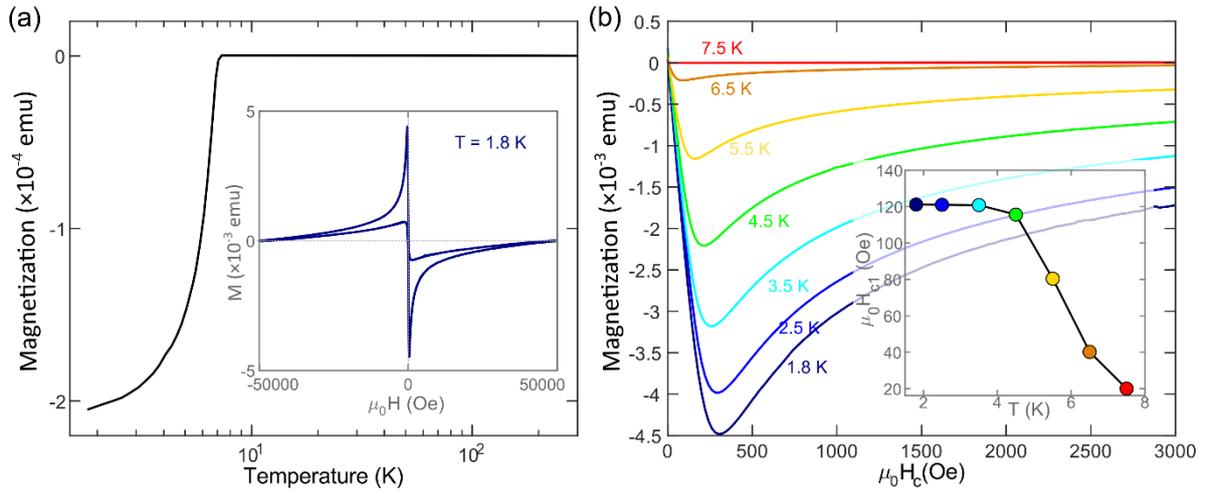

**Figure 2. Magnetization characterization of a NbSe$_2$ film on standard copy paper.** (a) Magnetization *vs*. temperature, measured after a zero-field cooling using 10 Oe of measuring field (sample 1). (Inset in a) Magnetization *vs*. magnetic field cycle, starting at zero-field, measured on a NbSe$_2$ film on paper after subtracting the paper contribution (b) Close-up of the first-magnetization curve at several temperatures. (Inset in b) $H_{c1}$ derived from the linear fits to the first-magnetization curve at low fields.

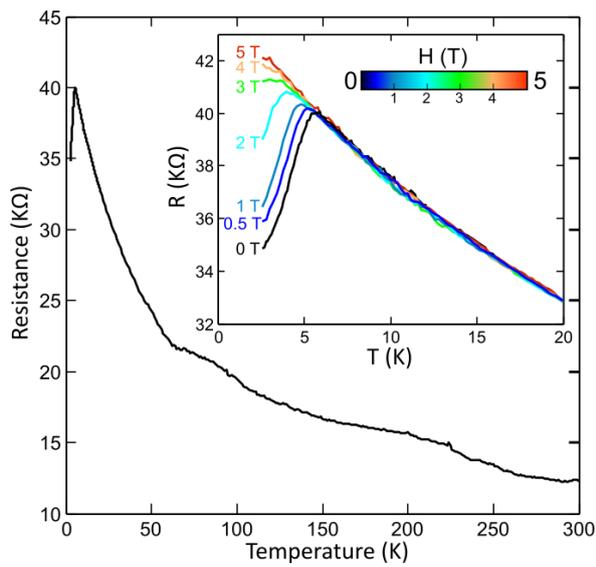

**Figure 3. Temperature dependent resistance of a NbSe$_2$ film on standard copy paper.** (a) Resistance *vs*. temperature measured on a NbSe$_2$-on-paper film at zero-field (sample 3, van der Pauw geometry). (Inset) Resistance *vs*. temperature curves measured at different applied magnetic fields.

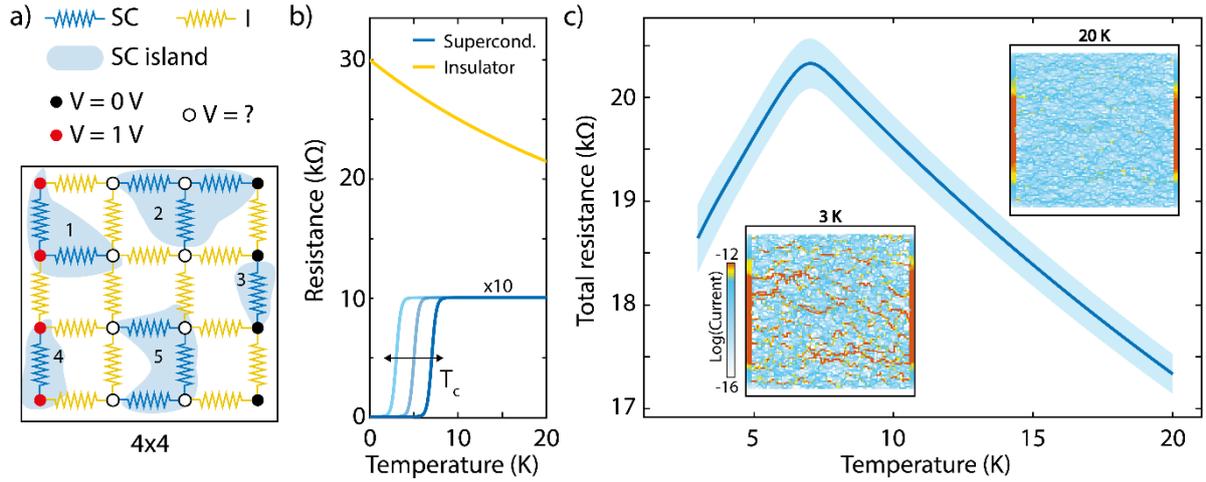

**Figure 4. Modelling the superconducting transition in a percolative film.** (a) Schematic of the random resistor network used to simulate the behavior of the interconnected network of NbSe$_2$ flakes. A small network of 4x4 resistors is shown as a simplified example. The network is composed by different resistors SC and I that represent the superconducting NbSe$_2$ flakes and the flake-to-flake hopping resistance respectively. (b) Temperature dependence of the SC and I resistors used for the simulation. We have considered a variation of the $T_c$ of the SC resistors from 1K to 7K to account for the thickness dependent $T_c$ of NbSe$_2$. The I resistors are modelled as a thermally activated hopping mechanism. (c) Mean total resistance calculated from 10 different 120x120 random resistors networks as a function of the temperature. The light blue bands correspond to a variation of 1 standard deviation from the mean curve. The insets show the current magnitude maps calculated in the normal (20K) and the superconducting (3K) state.

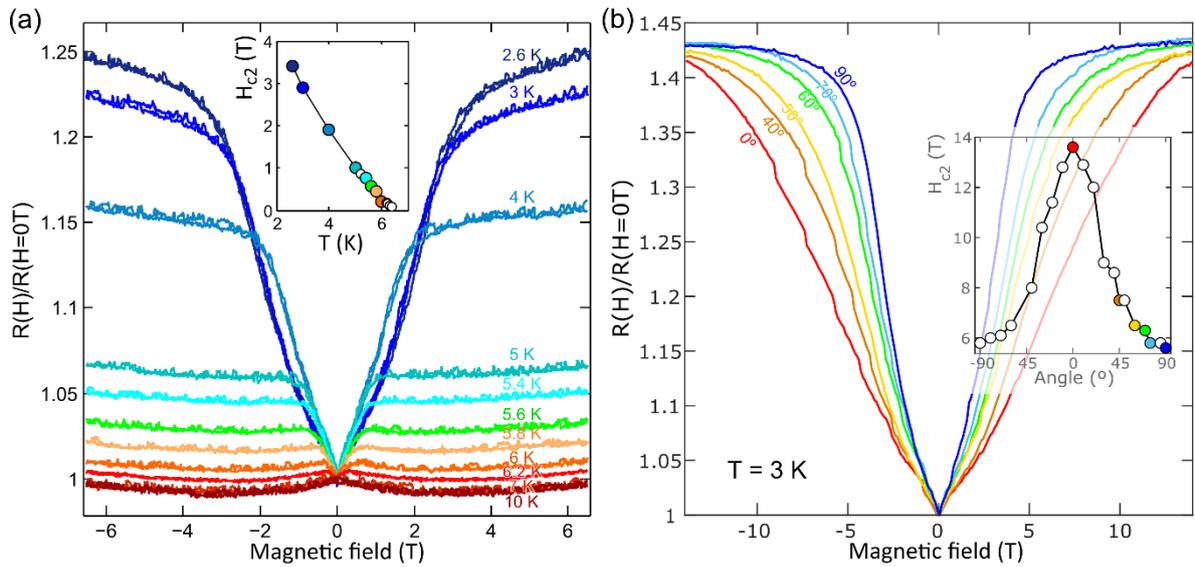

**Figure 5. Magneto-transport in a NbSe$_2$ film on paper at different temperatures and magnetic field rotation.** (a) Resistance *vs*. magnetic field, measured at different temperatures around the superconducting transition temperature (sample 4, four-terminal geometry). (Inset in a) Upper critical field, $H_{c2}$, values extracted for different temperatures from the magneto-transport data in (a). (b) Resistance of the NbSe$_2$ film vs. Magnetic field, measured for different magnetic field tilting angles ranging from: perpendicular (-90° and 90°) to parallel (0°) to the paper surface. (Inset in b) Ratio between the upper critical field $H_{c2}$ measured at a given magnetic field tilting angle (Θ) and that measured with the field perpendicular to the paper surface.

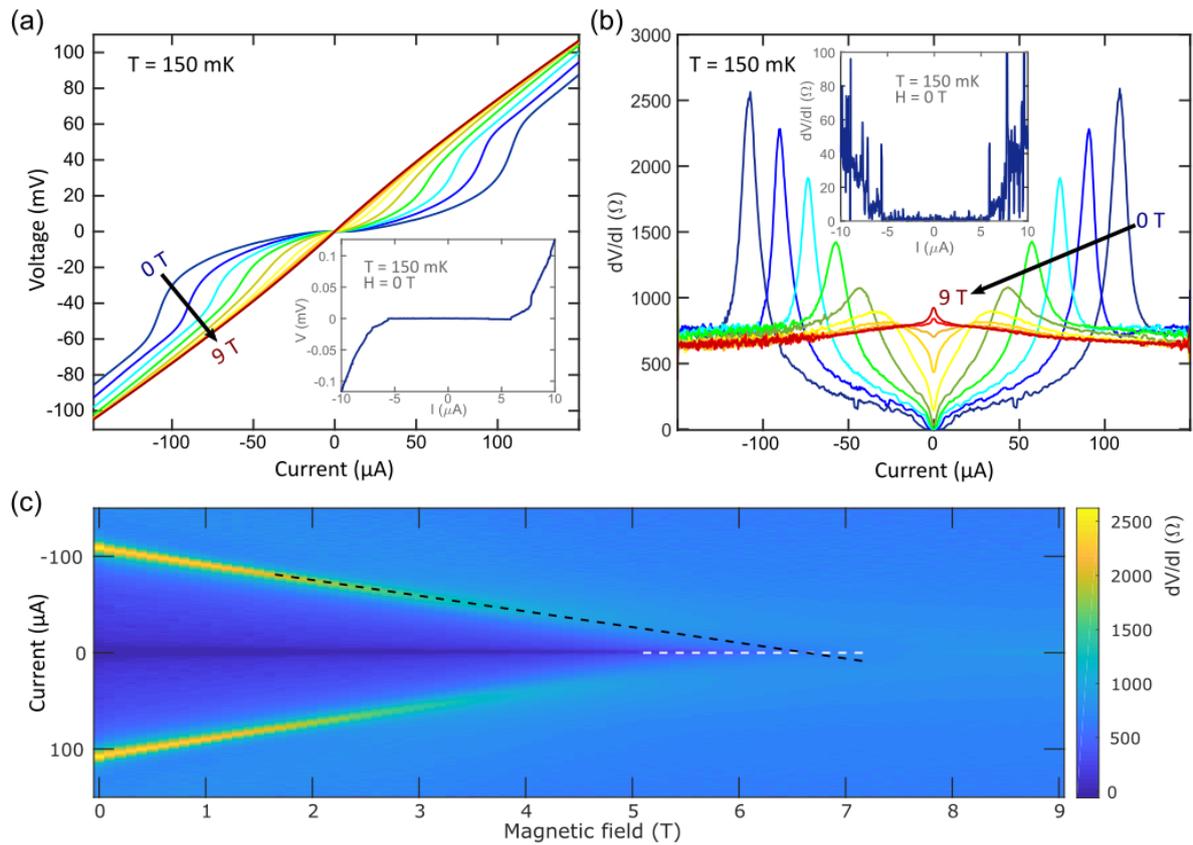

**Figure 6. Magneto-transport in a NbSe$_2$ film on paper at 150 mK.** (a) Voltage *vs.* current curves, measured at different magnetic fields (sample 5, van der Pauw geometry). (Inset in a) Detail of the voltage *vs.* current curve at $H$ = 0T which displays a zero-resistance state between ~±5 µA current bias. (b) Derivative of the voltage *vs.* current curves (d*V*/d*I*) at different fields (from 0 T to 9 T, in 1 T steps) where the destruction of the zero-resistance state at high magnetic field is evident. (c) False color map of the d*V*/d*I* (in the color axis) as a function of the bias current (vertical axis) and magnetic field (horizontal axis). The superconductivity is destroyed at $H$ ~ 6.5 T.

# Supporting Information

**Integrating superconducting van der Waals materials on paper substrates**

*Jon Azpeitia, Riccardo Frisenda\*, Martin Lee, Damian Bouwmeester, Wenliang Zhang, Federico Mompean, Herre S. J. van der Zant, Mar García-Hernández\*, Andres Castellanos-Gomez\**

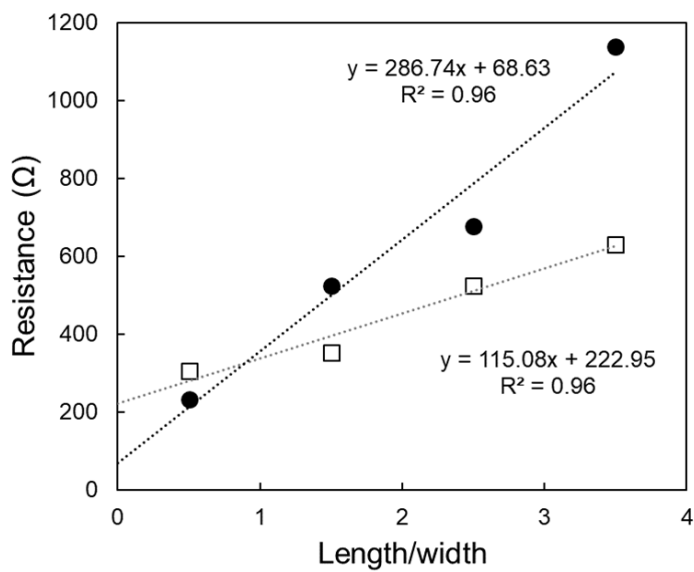

**Figure S1.** Transfer length characterization of two $NbSe_2$ films on paper to estimate the sheet resistance.

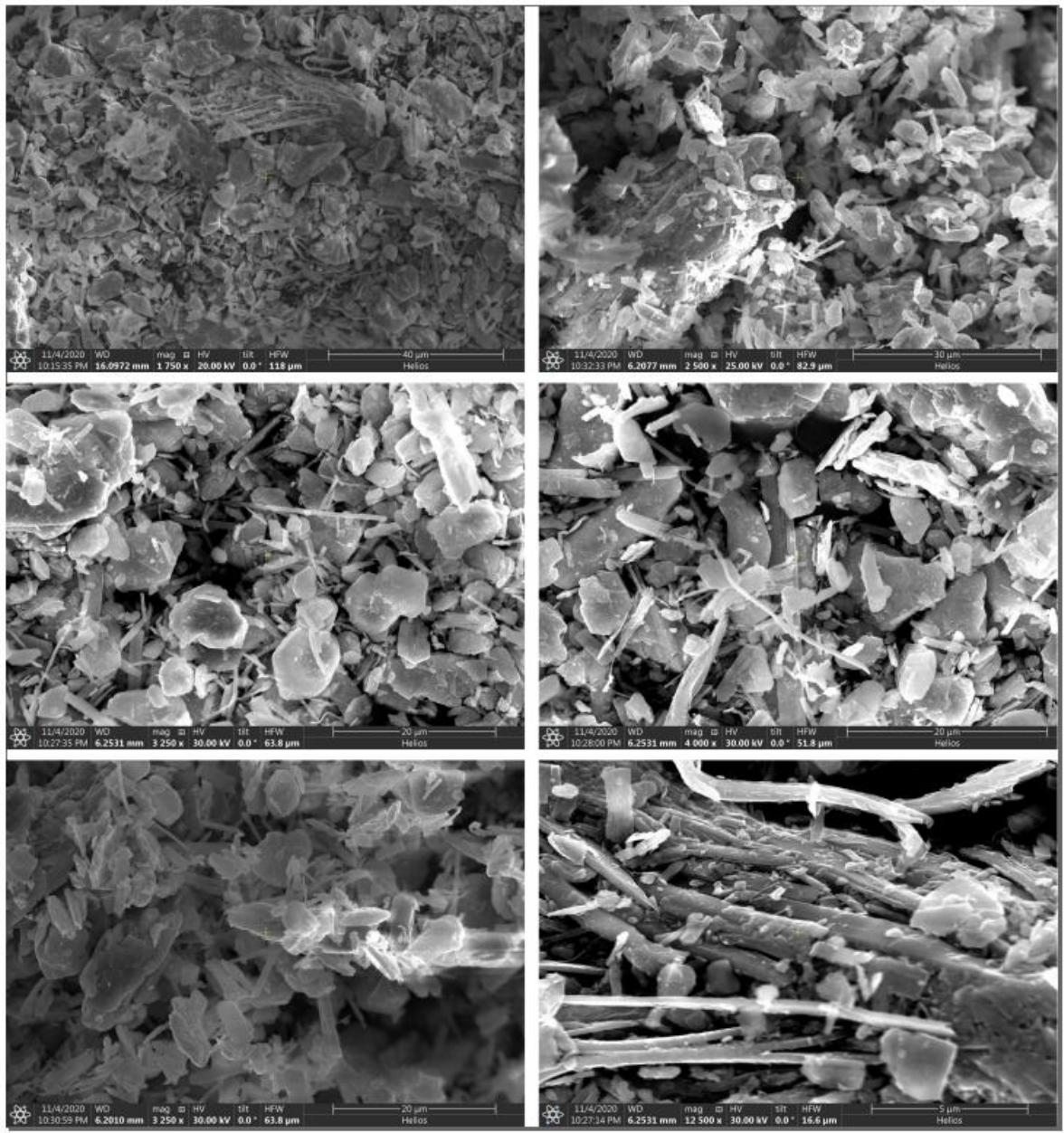

**Figure S2:** Scanning electron microscopy (SEM) images acquired on the NbSe$_2$ powder source material with different magnifications.

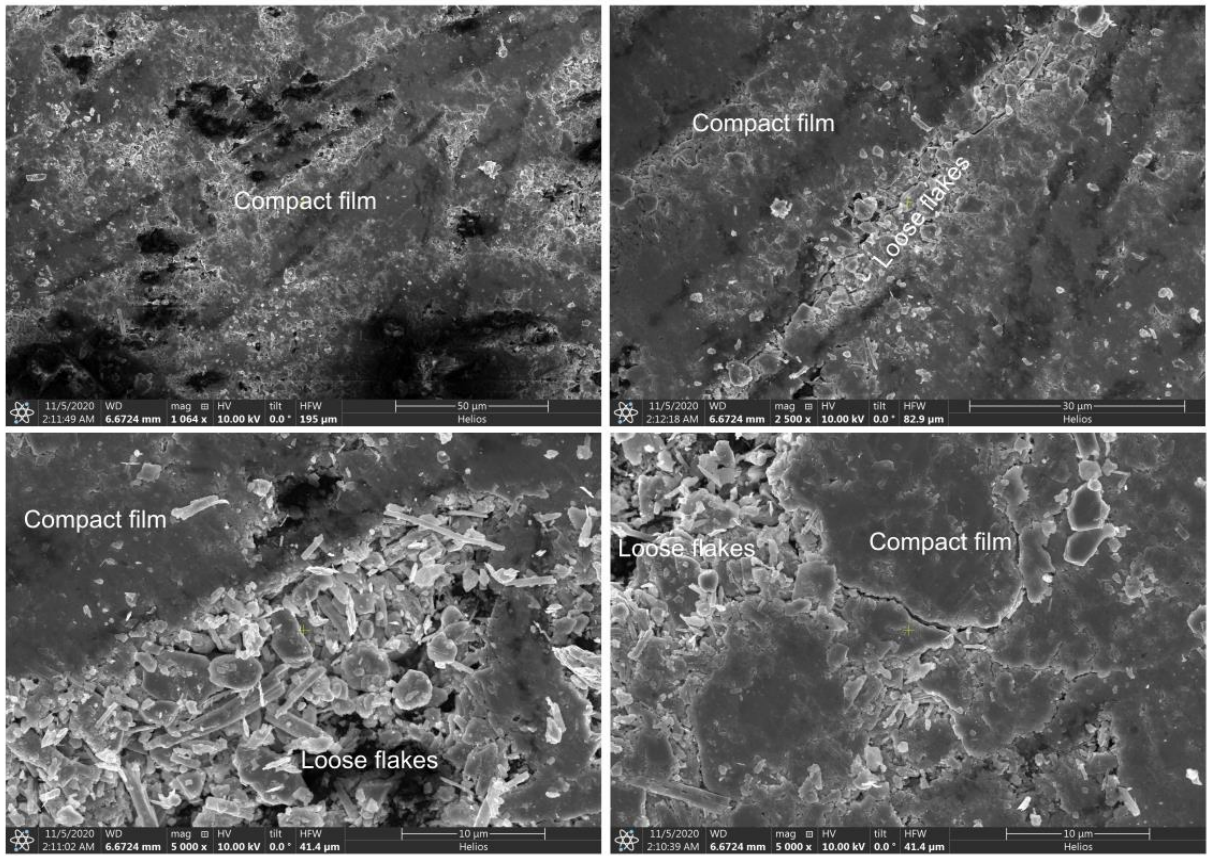

**Figure S3:** More scanning electron microscopy (SEM) images, at different magnifications, acquired on the NbSe$_2$ film on paper shown in Figure 1 of the main text. The overall film is formed by crushed NbSe$_2$ flakes that form a compact layer but in some of the gaps between paper fibers one can still observe loose NbSe$_2$ flakes.

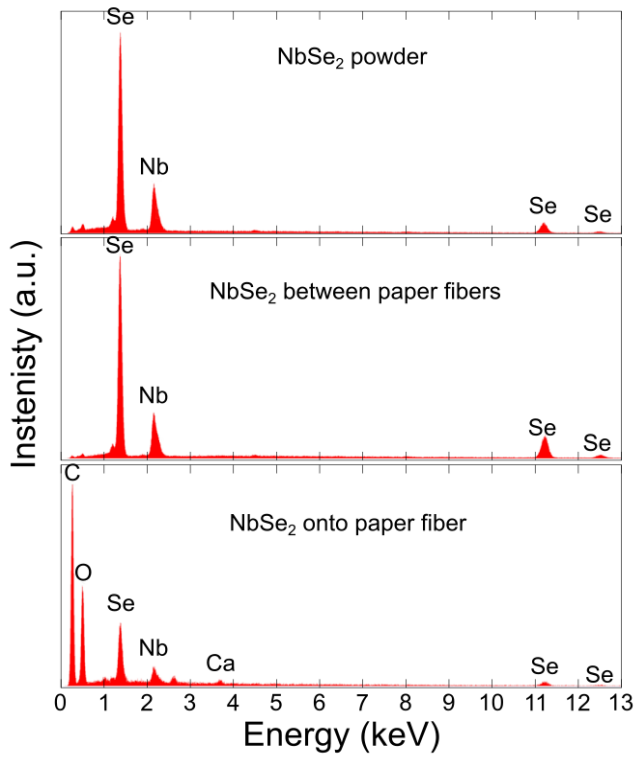

**Figure S4:** Energy dispersive X-ray spectroscopy (EDX) spectra shown in Figure 1d but in a broader energy range.

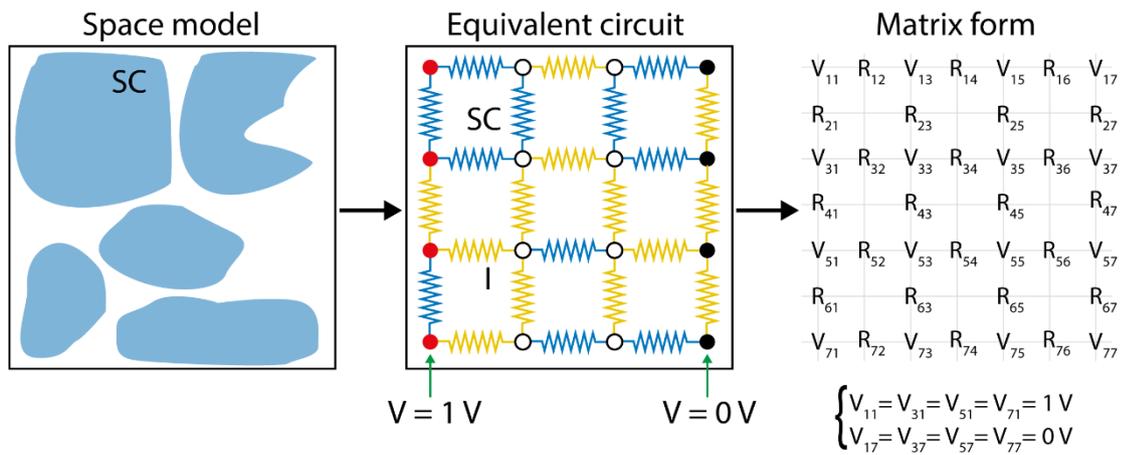

**Figure S5:** Illustration of the NbSe$_2$ percolative film as a network, (left) random network of superconducting particles, (center) random resistance network derived from the space model, (right) matrix form of the network used to solve the unknown voltages and the currents flowing through each resistor.

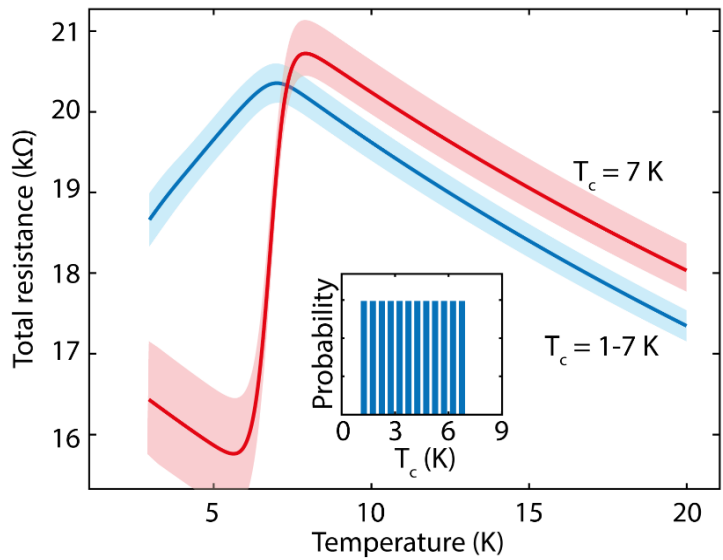

**Figure S6:** Mean total resistance of random resistors networks (size 120x120) as a function of the temperature with a uniformly distributed critical temperature between 1 K and 7 K (blue curve) and for a fixed critical temperature of 7 K (red). The colored bands correspond to a variation of 1 standard deviation from the mean curves.

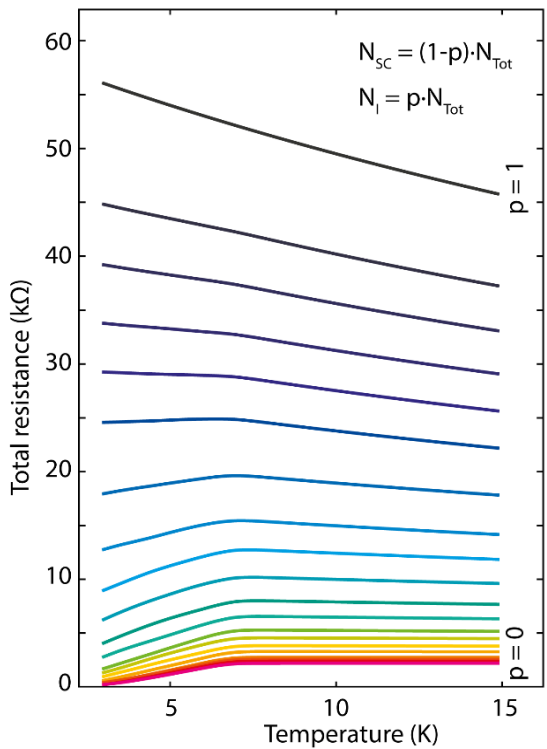

**Figure S7:** Total resistance of random resistors networks (size 120x120) as a function of the temperature and for different amount of superconducting elements in the film. Going from p=0 corresponding to a network entirely made of superconducting elements, to p=1 where the network is fully insulating. The network used in Figure 4 of the main text has p=0.65.

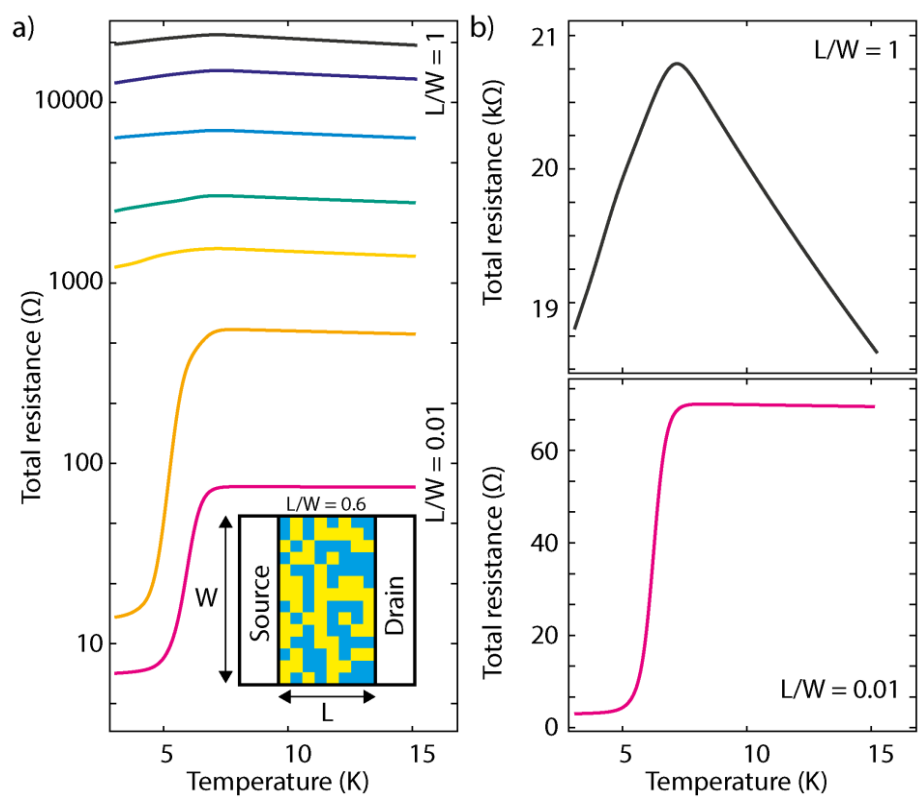

**Figure S8:** a) Semilogarithmic representation of the total resistance of random resistors networks (size 120x120, p = 0.65) as a function of the temperature calculated for different values of the aspect ratio L/W (where L is the channel length and W the width, see inset for an example with L/W = 0.6). b) Resistance versus temperature curves for a square channel (L/W = 1, top) and for a small aspect ratio channel (L/W = 0.01, bottom).